\documentclass[twocolumn,showkeys,aps,prb,showpacs]{revtex4-1}
\usepackage{graphicx}
\usepackage[CJKbookmarks,dvipdfm,colorlinks,linkcolor=blue,citecolor=blue]{hyperref}

\begin{document}

\title{Coexistence of intrinsic piezoelectricity and ferromagnetism induced by small biaxial  strain  in septuple-atomic-layer  $\mathrm{VSi_2P_4}$}

\author{San-Dong Guo$^{1}$, Wen-Qi Mu$^{1}$, Yu-Tong Zhu$^{1}$  and  Xing-Qiu Chen$^{2,3}$}
\affiliation{$^1$School of Electronic Engineering, Xi'an University of Posts and Telecommunications, Xi'an 710121, China}
\affiliation{$^2$Shenyang National Laboratory for Materials Science, Institute of Metal Research,
Chinese Academy of Science, 110016 Shenyang, Liaoning, P. R. China}
\affiliation{$^3$School of Materials Science and Engineering, University of Science and Technology of China,
Shenyang 110016, P. R. China}
\begin{abstract}
The septuple-atomic-layer  $\mathrm{VSi_2P_4}$ with the same structure of experimentally synthesized $\mathrm{MoSi_2N_4}$ is predicted to be a spin-gapless  semiconductor (SGS).  In this work,  the  biaxial  strain is applied to tune  electronic properties of  $\mathrm{VSi_2P_4}$, and it spans a wide range of properties upon the increasing strain from ferromagnetic metal (FMM) to SGS to ferromagnetic semiconductor (FMS) to SGS to ferromagnetic half-metal (FMHM). Due to broken inversion symmetry, the coexistence of
ferromagnetism and piezoelectricity can be achieved in  FMS $\mathrm{VSi_2P_4}$ with strain range of 0\% to 4\%. The calculated piezoelectric  strain coefficients $d_{11}$ for 1\%, 2\% and 3\% strains  are  4.61  pm/V, 4.94  pm/V and 5.27  pm/V, respectively, which are  greater than or close to  a typical
value of 5 pm/V for bulk piezoelectric materials.  Finally, similar to $\mathrm{VSi_2P_4}$, the coexistence of  piezoelectricity and ferromagnetism can be realized by strain in the $\mathrm{VSi_2N_4}$ monolayer. Our works show that  $\mathrm{VSi_2P_4}$ in FMS phase   with intrinsic piezoelectric properties  can have potential applications in spin electronic devices.

\end{abstract}

\pacs{71.20.-b, 77.65.-j, 72.15.Jf, 78.67.-n ~~~~~~~~~~~~~~~~~~~~~~~~~~~~~~~~~~~Email:sandongyuwang@163.com}

\maketitle

\section{Introduction}
Due to the advantages of high
speed, high integration density and high power transformers, the two-dimensional  (2D) magnetic materials have great potential applications for
nanoscale spintronic devices\cite{m1,m2,m3}. A few types of 2D magnetic materials have been
studied both in theory and in experiment\cite{m4,m5,m5-1,m5-2,m6,m7}.  For example, the $\mathrm{Cr_2Ge_2Te_6}$ is an intrinsic ferromagnetic 2D material\cite{m5}, and the monolayer  $\mathrm{VS_2}$ and $\mathrm{VSe_2}$ have also been experimentally demonstrated to be magnetic\cite{m5-1}.
The  $\mathrm{Mn_2C_6Se_{12}}$ and $\mathrm{Mn_2C_6S_6Se_6}$ monolayers are predicted to be Dirac SGSs with 100\% spin polarization, high Fermi velocities and  high
Curie temperatures\cite{m5-2}. Another particularly interesting  property of 2D materials is piezoelectricity,  which is used  for energy conversion
between electrical and mechanical energy.  It has been theoretically  reported that many kinds of 2D
materials have significant piezoelectric coefficients\cite{q7,q7-2,q7-3,q9,q10,q11,q12}. Experimentally discovered piezoelectricity  (such as $\mathrm{MoS_2}$\cite{q5,q6}, MoSSe\cite{q8}  and $\mathrm{In_2Se_3}$\cite{q8-1}) has  promoted the huge studies on piezoelectric properties of  2D materials.

It is  interesting and useful to combine the piezoelectricity and magnetism into the same kind of 2D material  for applications
in nanoscale spin electronic devices. The progress has been achieved in 2D  vanadium dichalcogenides, and  the  $\mathrm{VS_2}$,  $\mathrm{VSe_2}$  and Janus-VSSe are not only  magnetic semiconductors
but also exhibit appreciable  piezoelectricity\cite{m8}. Recently, the layered
2D $\mathrm{MoSi_2N_4}$ and  $\mathrm{WSi_2N_4}$ have been  synthesized by chemical vapor deposition (CVD)\cite{msn}.
And then a new kind of 2D family $\mathrm{MA_2Z_4}$ is proposed with $\alpha_i$ and  $\beta_i$ ($i$=1 to 6) phases  by intercalating $\mathrm{MoS_2}$-type  $\mathrm{MZ_2}$ monolayer into InSe-type  $\mathrm{A_2Z_2}$ monolayer\cite{msn,m20}.
Due to lacking inversion symmetry of $\mathrm{MA_2Z_4}$ with $\alpha_i$ phase,  the piezoelectricity can exist, such as  experimentally synthesized $\mathrm{MoSi_2N_4}$ and $\mathrm{WSi_2N_4}$ with $\alpha_1$ phase\cite{m21}. The $\alpha_1$-$\mathrm{VSi_2P_4}$  is predicted to be a SGS\cite{m20}, which may be easily tuned into FMS by strain. This will provide a platform to realize the coexistence of piezoelectricity and magnetism.

In this work,  the biaxial  strain effects on electronic  properties of  monolayer $\alpha_1$-$\mathrm{VSi_2P_4}$ are studied by the first principle calculations. In considered strain range, the  ferromagnetic (FM) ground state of $\mathrm{VSi_2P_4}$ is confirmed, and  it can change from FMM to SGS to FMS to SGS to FMHM with increasing strain. In FMS phase of $\mathrm{VSi_2P_4}$, the  piezoelectricities are investigated, and the calculated $d_{11}$ for 1\%, 2\% and 3\% strains  are  4.61  pm/V, 4.94  pm/V and 5.27  pm/V, respectively. Similar strain-induced phase transition can also  be achieved in monolayer  $\mathrm{VSi_2N_4}$, and only the critical point of phase transition is different from one of $\mathrm{VSi_2P_4}$.
 Our calculations show that the monolayer $\alpha_1$-$\mathrm{VSi_2P_4}$  may be  promising candidate for spintronic and piezoelectric applications by strain engineering.

The rest of the paper is organized as follows. In the next
section, we shall give our computational details and methods.
 In the third section,  we shall present main results of septuple-atomic-layer  $\mathrm{VSi_2P_4}$. Finally, we shall give our discussion and conclusions in the fourth section.

\section{Computational detail}
Calculations are based on spin-polarized density functional
theory (DFT)\cite{1} using the popular  generalized gradient
approximation of Perdew, Burke and  Ernzerhof  (GGA-PBE)\cite{pbe} as the exchange-correlation  functional.
Projector augmented wave (PAW) potentials are used
in all calculations,  as implemented
in the plane-wave code VASP\cite{pv1,pv2,pv3}. A vacuum spacing of more than 32 $\mathrm{{\AA}}$ is used in the direction normal to the 2D monolayer in order to avoid interactions between two neighboring images.
The kinetic energy cutoff is set to 500 eV, and the total energy  convergence criterion is set
to $10^{-8}$ eV.  The ionic relaxation is performed until the
force on each atom is less than 0.0001 $\mathrm{eV.{\AA}^{-1}}$.
Piezoelectricity is studied by  using density functional perturbation theory (DFPT)\cite{pv6}.
The reciprocal space is represented by the Monkhorst-Pack special
k-point scheme with 15$\times$15$\times$1 meshes for the  calculations of electronic structure  and elastic coefficients $C_{ij}$ , and 9$\times$16$\times$1 grid meshes
for  the energy of different magnetic configurations and piezoelectric   stress  coefficients $e_{ij}$.
The 2D elastic coefficients $C^{2D}_{ij}$
 and   piezoelectric stress coefficients $e^{2D}_{ij}$
have been renormalized by   $C^{2D}_{ij}$=$Lz$$C^{3D}_{ij}$ and $e^{2D}_{ij}$=$Lz$$e^{3D}_{ij}$, where the $Lz$ is  the length of unit cell along z direction.

\begin{figure}
  \includegraphics[width=7.0cm]{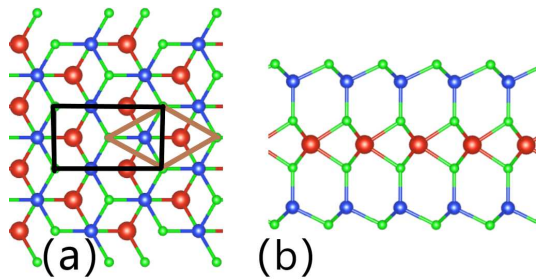}
  \caption{(Color online)The (a) top view and (b) side view of  crystal structure of monolayer  $\mathrm{VSi_2P_4}$. The large red balls represent V atoms, and the middle blue balls for Si atoms, and the  small green balls for P atoms. The primitive cell and rectangle supercell
   are marked by brown and black lines, respectively.}\label{t0}
\end{figure}
\begin{figure}
  \includegraphics[width=8cm]{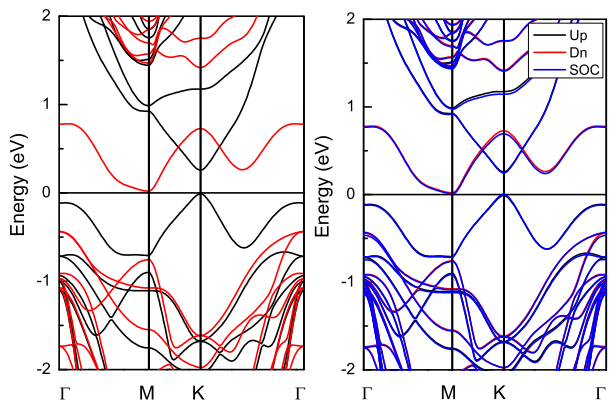}
  \caption{(Color online)The  energy band structures of unstrained $\mathrm{VSi_2P_4}$, include using  GGA (Left) or  GGA+SOC and GGA (Right) with FM state.}\label{band}
\end{figure}
\begin{figure}
  \includegraphics[width=8cm]{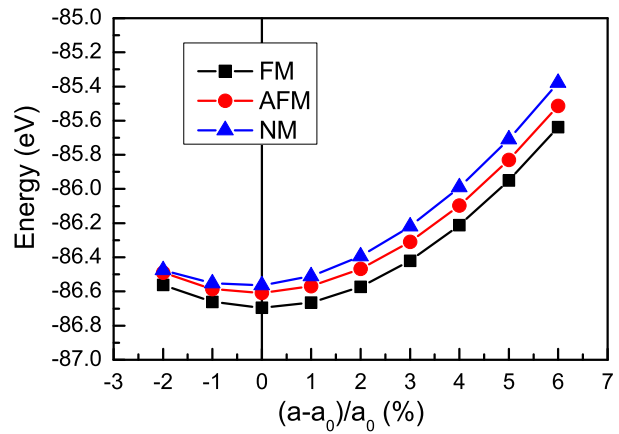}
  \caption{(Color online) Calculated energy of   FM state,   AFM state and NM state as a function of strain with rectangle supercell.  }\label{energy}
\end{figure}

  \begin{figure*}
  \includegraphics[width=14cm]{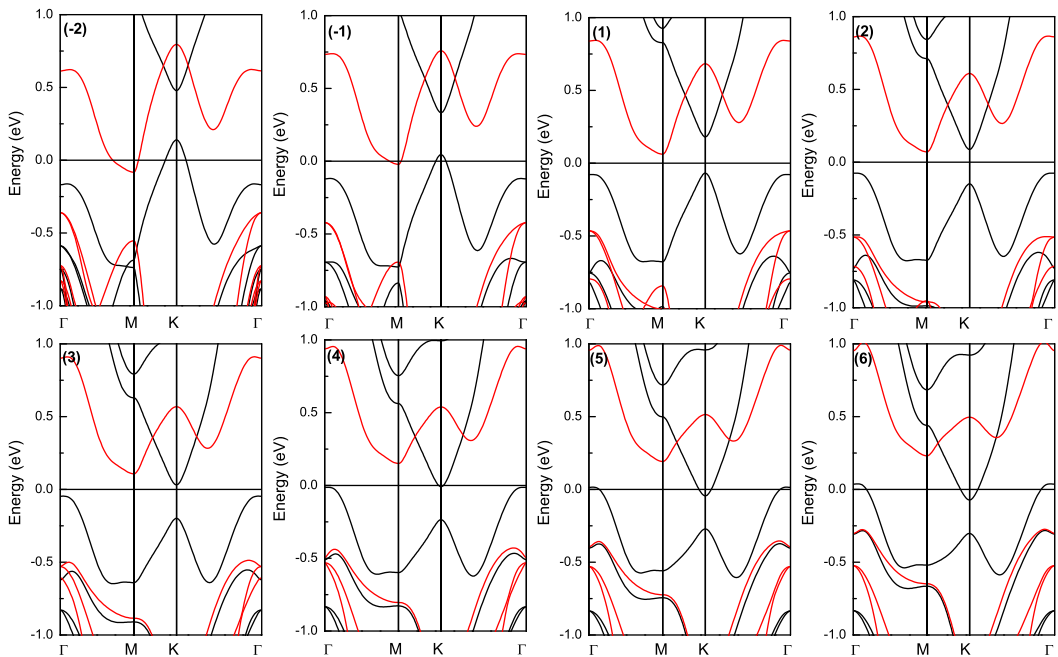}
  \caption{(Color online) The  energy band structures of FM $\mathrm{VSi_2P_4}$ using GGA with strain from -2\% to 6\% except 0\%, which has been shown in \autoref{band}. The black and red lines represent the spin-up and spin-down bands.}\label{band1}
\end{figure*}
  \begin{figure}
  \includegraphics[width=8cm]{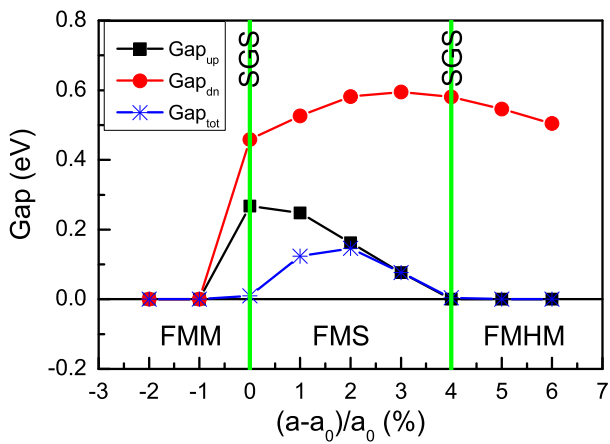}
  \caption{(Color online)The majority-spin gap ($\mathrm{Gap_{up}}$), the minority-spin gap ($\mathrm{Gap_{dn}}$) and the total gap ($\mathrm{Gap_{tot}}$) of FM $\mathrm{VSi_2P_4}$ as a function of  strain using GGA. The electronic properties change  from FMM to SGS to FMS to SGS to FMHM with the increasing strain.   }\label{gap}
\end{figure}
\begin{figure}
  \includegraphics[width=8cm]{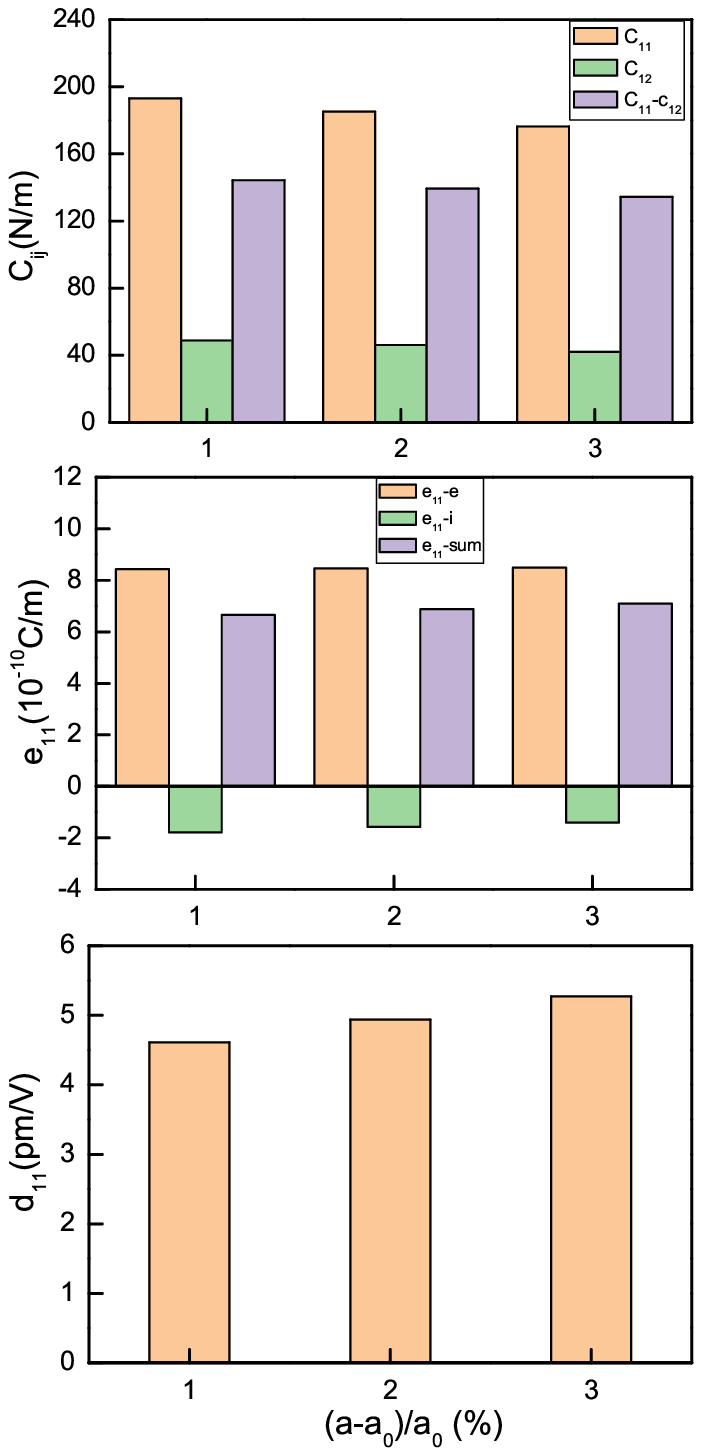}
  \caption{(Color online)For monolayer $\mathrm{VSi_2P_4}$ , (Top) the elastic constants  $C_{ij}$, (Middle) piezoelectric stress coefficients  $e_{11}$ along with the ionic contribution and electronic contribution, and (Bottom) piezoelectric strain coefficients  $d_{11}$  with the  strain being 1\%, 2\% and 3\%.}\label{ced}
\end{figure}

\section{Main calculated results}
The $\alpha_1$-$\mathrm{VSi_2P_4}$ is predicted to be  a parabolic FM  SGS with the total magnetic moment 1.0 $\mu_B$\cite{m20}.
Firstly, we  relax the geometric structures of  $\mathrm{VSi_2P_4}$ with spin-polarized calculations, which is plotted in \autoref{t0}.
The seven atomic layers of N-Si-N-V-N-Si-N  are observed  with a $\mathrm{VN_2}$ layer sandwiched between two Si-N bilayers.
The optimized lattice constants is 3.486 $\mathrm{{\AA}}$, which is in good agreement with the
previous theoretical result\cite{m20}. We calculate electronic band
structures of  $\mathrm{VSi_2P_4}$ using GGA and GGA plus spin orbital coupling (SOC), which are plotted in \autoref{band}. It is clearly seen that  the Fermi energy level happens to touch the minority-spin conduction band minimum (CBM) at M point  and the majority-spin valence band maximum (VBM)  at K point at the same time, and  a SGS is achieved. Calculated results show that the SOC has little effects on  energy bands of  $\mathrm{VSi_2P_4}$, so we use GGA to investigate the role of strain on the electronic structures of $\mathrm{VSi_2P_4}$.

Strain is a very effective way to tune the electronic structures, topological properties, transport and   piezoelectric  properties of 2D materials\cite{m12,m13,m14,m15}.
 The biaxial strain can be simulated by  $(a-a_0)/a_0$ with  $a$ and $a_0$ being  the strained and  unstrained lattice constants.  To determine the ground state of strained   $\mathrm{VSi_2P_4}$,a rectangle supercell (in \autoref{t0}) is used  to construct  two different initial magnetic configurations, including antiferromagnetic (AFM) and FM states.  The energy of   FM state,   AFM state and non-magnetic (NM) state as a function of strain are shown in \autoref{energy}. It is clearly seen that $\mathrm{VSi_2P_4}$  prefers FM ground state in considered strain range, and the energy difference between AFM and FM states increases  from  73 meV to 124 meV, when the strain changes from -2\% to 6\%. This means that the applied biaxial strain
 can effectively enhance  the magnetic coupling of $\mathrm{VSi_2P_4}$, increasing the
Curie temperature.

The  energy band structures of FM $\mathrm{VSi_2P_4}$ using GGA with strain from -2\% to 6\% except 0\% are shown in \autoref{band1}, and the majority-spin,  minority-spin  and total gaps  are plotted in \autoref{gap}. From -2\% to 6\%,  the first conduction band of majority-spin moves toward the lower energy, and touches exactly the Fermi level at 4\% strain, and then crosses the the Fermi level at  $>$4\% strain. The similar trend can be observed for the first conduction band of minority-spin, when the strain changes from 6\% to -2\%,  and happen to  touch the Fermi level at 0\% strain.
For the first valence band of majority-spin, the energy eigenstates around the K point vary toward lower energy with strain from -2\% to 6\%, and touch the Fermi level at 0\% strain, while the energy eigenstates around the $\Gamma$ point move toward higher energy, and contact the Fermi level at 4\% strain. These lead to  diverse properties of $\mathrm{VSi_2P_4}$ upon strain  from FMM to SGS to FMS to SGS to FMHM. The compressive strain makes both majority-spin and  minority-spin gaps be zero, and a FMM is achieved.
When the tensile strain is less than 4\%, both majority-spin and  minority-spin gaps  are nonzero, and then  the total gap  is positive value.  In this strain range, a FMS can be induced by tensile strain. With tensile strain being larger than 4\%,  only minority-spin gap is nonzero, and the majority-spin gap become zero, which gives rise to a FMHM. At the critical states of 0\% and 4\%, the SGS can be observed, but they are different.
At 0\% strain, there is a gap  for both the majority and minority channels, while there is
no gap between the majority channel in the valence band and the minority channel in the conduction band.
At 4\% strain, the majority channel is gapless,  while the minority channel is separated  by
a gap. Similar stress-tuned SGS can be observed in  the ferromagnetic
semiconductor  $\mathrm{HgCr_2Se_4}$\cite{jpcm}.

\begin{figure}
  \includegraphics[width=8cm]{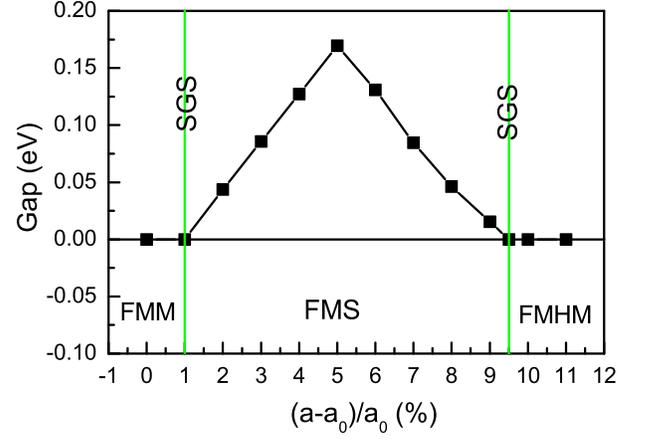}
  \caption{(Color online)The total gap ($\mathrm{Gap}$) of FM $\mathrm{VSi_2N_4}$ as a function of  strain using GGA. The electronic properties change  from FMM to SGS to FMS to SGS to FMHM with the increasing strain.   }\label{gap1}
\end{figure}

The $\mathrm{VSi_2P_4}$ has the $\bar{6}m2$ point group, and then lacks the inversion symmetry, which makes $\mathrm{VSi_2P_4}$ to be piezoelectric.
A piezoelectric material  should  be a semiconductor to  prohibit current leakage. It is interesting to investigate the piezoelectric properties of  $\mathrm{VSi_2P_4}$ with the strain range of 0\% to 4\%, because the coexistence of piezoelectricity and magnetism  can be achieved in the  same kind of material. The piezoelectric stress tensors  $e_{ijk}$ and strain tensor $d_{ijk}$  can be used to characterize the linear piezoelectric effect of monolayer $\mathrm{VSi_2P_4}$, which include  ionic
and electronic contributions.  Using the Voigt
notation, the  $e_{ijk}$ and $d_{ijk}$ can be represented as  $e_{ij}$ and  $d_{ij}$, respectively. The $e_{ik}$ is connected  with $d_{ij}$ by:
\begin{equation}\label{aaaaa}
    e_{ik}=d_{ij}C_{jk}
\end{equation}
For 2D semiconductors, in general,
in-plane  stresses and strains are only allowed,
while the out-of-plane is strain/stress free\cite{q7,q9,q11}. For $\bar{6}m2$ point group,  the  $C_{ij}$, $e_{ij}$ and $d_{ij}$  of   monolayer  $\mathrm{VSi_2P_4}$ become:
\begin{equation}\label{pe1}
  \left(
    \begin{array}{ccc}
      e_{11} &-e_{11} & 0 \\
    0 &0 & -e_{11}\\
      0 & 0 & 0 \\
    \end{array}
  \right)
  \end{equation}
  \begin{equation}\label{pe1}
  \left(
    \begin{array}{ccc}
        d_{11} & -d_{11} & 0 \\
    0 &0 & -2d_{11} \\
      0 & 0 & 0 \\
    \end{array}
  \right)
  \end{equation}
  \begin{equation}\label{pe1}
    \left(
    \begin{array}{ccc}
      C_{11} & C_{12} &0 \\
     C_{12} & C_{11} &0 \\
     0 & 0 & \frac{C_{11}-C_{12}}{2} \\
    \end{array}
  \right)
   \end{equation}
Solving the about equations,  the only  independent in-plane $d_{11}$ is:
\begin{equation}\label{pe2-7}
    d_{11}=\frac{e_{11}}{C_{11}-C_{12}}
\end{equation}

A rectangle supercell is used for the calculation of
$e_{ij}$, as shown in \autoref{t0}. The $C_{ij}$ are calculated by using strain-stress relationship.
The elastic constants  $C_{ij}$,  piezoelectric stress coefficients  $e_{11}$ along with the ionic contribution and electronic contribution, and piezoelectric strain coefficients  $d_{11}$  of  monolayer $\mathrm{VSi_2P_4}$  with the  strain being 1\%, 2\% and 3\% are shown in \autoref{ced}.
All calculated elastic coefficients of  $\mathrm{VSi_2P_4}$  with the  strain being 1\%, 2\% and 3\% satisfy the Born stability criteria\cite{ela}. It is found that the ionic and electronic parts have opposite contributions to $e_{11}$. The increasing strain can enhance the $e_{11}$, and  these values for 1\%, 2\% and 3\% strains  are 6.65$\times$$10^{-10}$ C/m,     6.88$\times$$10^{-10}$ C/m and   7.09$\times$$10^{-10}$ C/m, respectively. The trend is mainly due to
decreased ionic contribution (absolute value). The corresponding values of $d_{11}$ for 1\%, 2\% and 3\% strains  are  4.61  pm/V, 4.94  pm/V and 5.27  pm/V, respectively.  The strain-improved $d_{11}$ is due to the enhanced $e_{11}$ and reduced $C_{11}$-$C_{12}$ according to \autoref{pe2-7}.
These results are comparable with or larger than ones of  TMD monolayers such as  $\mathrm{MoS_2}$
obtained from DFT calculations\cite{q9,q11}. The values of $d_{11}$ of FMS $\mathrm{VSi_2P_4}$ are  greater than or close to  a typical
value for bulk piezoelectric materials, about 5 pm/V\cite{q9}.  This indicates that FMS $\mathrm{VSi_2P_4}$ caused by small strain show potential in  piezoelectric devices.

The  $\mathrm{VSi_2N_4}$ has the same crystal structure with  $\mathrm{VSi_2P_4}$, and is predicted to be a FMM with  the total magnetic moment 0.97 $\mu_B$\cite{m20}, which is very close to 1.0 $\mu_B$. This makes us believe  that small strain may induce FMM to FMS transition. Firstly, the ground state of $\mathrm{VSi_2N_4}$ is determined to be  FM in the strain range of 0\% to 11\% by comparing the energy of FM,  AFM  and NM, and their energy bands are calculated. The total gap of FM $\mathrm{VSi_2N_4}$ as a function of  strain using GGA is plotted in \autoref{gap1}. It is found that only about 1\% strain can make $\mathrm{VSi_2N_4}$ become FMS, which can be maintained in the strain range of 1\% to 9.5\%.
When the strain increases, the electronic properties of $\mathrm{VSi_2N_4}$ vary  from FMM to SGS to FMS to SGS to FMHM, which is the same with $\mathrm{VSi_2P_4}$. The piezoelectric properties of  $\mathrm{VSi_2N_4}$ at 3\% strain are studied as a representative. The calculated $C_{11}$ and $C_{12}$ are   431.17 N/m  and 117.61 N/m, respectively, and the $e_{11}$ is 6.71$\times$$10^{-10}$ C/m. According to \autoref{pe2-7}, the $d_{11}$ can be attained for 2.14 pm/V.

\section{Discussion and Conclusion}
It is well known that GGA underestimates semiconductor gap, and the  hybrid functional HSE06 may give a more appropriate gap. 
However, for $\mathrm{MoSi_2N_4}$ monolayer, the HSE06  overestimates it's gap. The calculations show that $\mathrm{MoSi_2N_4}$ monolayer is  an indirect gap semiconductor  with the gap of 1.744 eV (GGA) or 2.297 eV (HSE06), and the experimental value is 1.94 eV\cite{msn}. It is found that the GGA gap is more closer to the experimental value than HSE06 one. So, it may be more suitable for $\mathrm{VSi_2P_4}$ to use GGA to study it's electronic properties. For  $\mathrm{VSi_2P_4}$ monolayer, the GGA may underestimates it's gap, but our conclusion should be  qualitatively correct, and only the critical points of phase transition change.

We have demonstrated that strain can effectively tune the electronic properties of  $\mathrm{VSi_2P_4}$ monolayer through first-principles simulations.
The results show that FM state in considered strain range  is more energetically preferred than AFM and NM states for  $\mathrm{VSi_2P_4}$ monolayer.
With the increasing strain, the electronic properties of $\mathrm{VSi_2P_4}$ vary  from FMM to SGS to FMS to SGS to FMHM. The corresponding values
of $d_{11}$ of $\mathrm{VSi_2P_4}$ in FMS phase for 1\%, 2\% and 3\% strains are   4.61  pm/V, 4.94  pm/V and 5.27  pm/V, respectively, which are greater than 5 pm/V, a typical value for bulk piezoelectric materials. Similar strain dependence of electronic properties can also be achieved in monolayer  $\mathrm{VSi_2N_4}$, and the piezoelectricity and ferromagnetism can also  coexist by strain tuning. The  $\mathrm{VSi_2P_4}$ and  $\mathrm{VSi_2N_4}$ may be promising 2D materials  for applications in nanoscale spin electronic devices due to
the combination of  piezoelectric and magnetic properties.

\begin{acknowledgments}
This work is supported by the Natural Science Foundation of Shaanxi Provincial Department of Education (19JK0809). We are grateful to the Advanced Analysis and Computation Center of China University of Mining and Technology (CUMT) for the award of CPU hours and WIEN2k/VASP software to accomplish this work.
\end{acknowledgments}

\end{document}